\documentclass[a4paper]{article}

\usepackage{INTERSPEECH2020}
\usepackage{subfigure}
\usepackage{booktabs} 
\usepackage{array}
\usepackage{amsmath,graphicx}
\usepackage{multirow}
\usepackage{hyperref}

\title{ Prosody Learning Mechanism for Speech Synthesis System \\Without Text Length Limit}

\name{Zhen Zeng, Jianzong Wang$^*$\thanks{ $^*$Corresponding author: Jianzong Wang, jzwang@188.com }, Ning Cheng, Jing Xiao}
\address{
  Ping An Technology (Shenzhen) Co., Ltd. }
\email{ \{zengzhen830, wangjianzong347, chengning211, xiaojing661\}@pingan.com.cn}

\begin{document}

\maketitle
\begin{abstract}
 
  Recent neural speech synthesis systems have gradually 
  focused on the control of prosody to improve the quality 
  of synthesized speech, but they rarely consider the 
  variability of prosody and the correlation between prosody 
  and semantics together. In this paper, a prosody learning 
  mechanism is proposed to model the prosody of speech based 
  on TTS system, where the prosody information of speech is 
  extracted from the mel-spectrum by a prosody learner and 
  combined with the phoneme sequence to reconstruct the 
  mel-spectrum. Meanwhile, the sematic features of text from 
  the pre-trained language model is introduced to improve the 
  prosody prediction results. In addition, a novel self-attention 
  structure, named as local attention, is proposed to lift 
  this restriction of input text length, where the relative 
  position information of the sequence is modeled by the 
  relative position matrices so that the position encodings 
  is no longer needed. Experiments on English and Mandarin show 
  that speech with more satisfactory prosody has obtained 
  in our model. Especially in Mandarin synthesis, 
  our proposed model outperforms baseline model with a MOS gap 
  of 0.08, and the overall naturalness of the synthesized 
  speech has been significantly improved.

\end{abstract}

\noindent\textbf{Index Terms}: speech synthesis, text-to-speech, self-attention, prosody modeling

\section{Introduction}

Speech synthesis, also known as text-to-speech (TTS), 
has attracted a lot of attention and obtained satisfactory 
results in recent years due to the advances in deep learning. 
Several TTS systems based on deep networks were proposed, 
such as Char2Wav \cite{Char2wav}, Tacotron2 \cite{Tacotron2}, 
DeepVoice3 \cite{DeepVoice3}, Transformer TTS \cite{TransformerTTS}, 
FastSpeech \cite{FastSpeech} and ParaNet \cite{ParaNet}. 
These systems usually first predict the acoustic feature sequence 
from the input text sequence, and then generate waveform from 
the acoustic feature sequence using vocoder such as 
Griffin-Lim \cite{GriffinLim}, WaveNet \cite{wavenet}, 
WaveRNN \cite{WaveRNN}, WaveGlow \cite{WaveGlow} 
and GAN-TTS \cite{GAN-TTS}.

Although current speech synthesis systems have obtained 
high-quality speech, it is difficult for them to achieve 
satisfactory results in long text speech synthesis scenarios. 
In the sequence-to-sequence TTS model, since the monotonicity 
and locality properties of TTS alignment are not fully utilized, 
the alignment procedure lacks robustness in inference, 
which leads to skipping or repeating words, incomplete 
synthesis, or an inability to synthesize long utterances 
\cite{Location-relative-attention-TTS}. To address the issue, 
many monotonic attention mechanisms are presented 
\cite{ForwardAttetnion, StepwiseAttention}, where only 
the alignment paths satisfying the monotonic condition 
are taken into consideration at each decoder timestep. 
In \cite{Location-relative-attention-TTS, RepresentationMixTTS}, 
the location-based GMM attention introduced in \cite{GMM-Attention} 
is also studied in TTS systems to generalize to long utterances. 
Especially, AlignTTS \cite{AlignTTS} proposes an alignment loss 
to model the alignment between text and mel-spectrum, and uses 
a length regulator to adjust the alignment, which solves the 
instability problem of the alignment and is very efficient. 
However, since the self-attention of Transformer 
\cite{AttentionAllYouNeed} is used to model the dependencies 
of input sequence elements in AlignTTS, the positional encodings 
are required to introduce the positional information, 
which limits the maximum length of input text. 
In this paper, a novel self-attention mechanism is proposed
to remove the need for the positional encodings and 
lift the restriction of input text length.



On the other hand, the prosody of speech directly affects 
the overall listening perception of the voice, especially for 
long utterances. In order to improve the naturalness of 
synthetic speech, it is necessary for TTS systems to model 
prosody information. In \cite{RobustFineGrainedProsody}, 
a prosody embedding is introduced for emotional and 
expressive speech synthesis, which enables fine-grained control 
of the speaking style. In \cite{Fully-hierarchical-Prosody}, 
an interpretable latent variable model for prosody based on 
Tacotron 2 is presented to model phoneme-level and word-level 
prosody information of speech. \cite{Prosody-Prior} proposes 
a quantized latent feature for the prosody of speech, and trains 
an autoregressive prior model to generate natural samples 
without a reference speech. These prosody control methods 
enable us to learn the prosody from speech and fine-grained 
the synthesized speech, but they still cannot effectively 
predict the correct prosody according to the input text.  
One reason is that the prosody information of speech 
generally depends on the meaning of text, while only the 
phoneme information of text is used as the input in current 
mainstream TTS systems, which limits the capabilities of 
modeling the prosody of speech. 
In \cite{Bert-TTS1,Bert-TTS2,Bert-TTS-Mandarin}, 
the textual knowledge from BERT \cite{Bert} is introduced 
into TTS systems to improve the prosody of speech, 
but they ignore the variability of prosody. 
For example, the same text may produce speech with 
different prosody due to pronunciation uncertainty.

In this works, we propose a novel self-attention mechanism, 
named as local attention, to model the timing dependencies, 
which abandons positional encoding and uses a relative 
position matrix to model the influence of the positional 
relationship of input sequence. At the same time, 
we introduce the prosody learning mechanism for feed-forward 
TTS systems, where a prosody embedding for each phoneme is 
learned from the mel-spectrum in training. In addition, 
a prosody predictor is designed to predict the prosody 
embedding according to text and phoneme, where a pre-trained 
language model is applied to introduce the meaning of text. 
And the main contributions of our works as follows:
\begin{itemize}
  \item The local attention is proposed to model sequence 
  dependency so that the speech synthesis system is 
  no longer restricted by the length of the input text sequence;
  \item The prosody learning mechanism is designed for 
  feed-forward TTS system to capture the prosody information 
  from the mel-spectrum. Thanks to the separation of the prosody 
  information, more accurate mel-spectrum prediction model is obtained;
  \item In order to improve the prosody prediction results, 
  the pre-trained language model is applied in TTS system to 
  introduce the sematic features of text.
\end{itemize}

\section{Our Methods} 

In this section, we propose the local attention that 
can capture the sequence dependency without the positional 
encoding, and the prosody learning mechanism that can 
separate the prosody of speech from the pronunciation of phoneme. 
And based on the architecture of AlignTTS, a feed-forward speech 
synthesis system with prosody learning ability is designed, 
which has not limit on the length of input text.

\begin{figure*}[t]
  \centering
  \subfigure[ Model Training with Prosody Learning Mechanism] { 
    \begin{minipage}[b]{0.6\linewidth}
      \centering
      \includegraphics[width=0.902\linewidth]{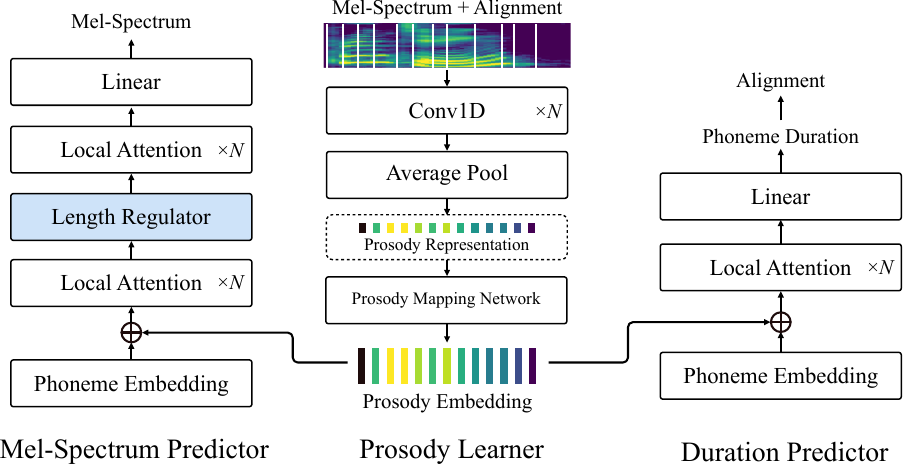}
    \end{minipage}
    \label{figure-prosody-modeling}
  }
  \subfigure[ Prosody Predictor ] {
    \begin{minipage}[b]{0.25\linewidth}
      \centering
      \includegraphics[width=0.671\linewidth]{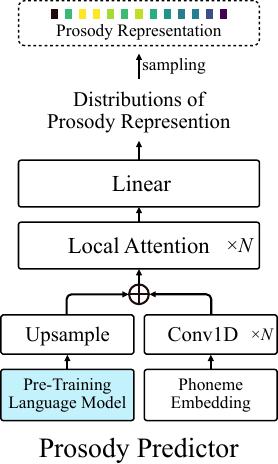}
    \end{minipage}
    \label{figure-prosody-predictor}
  }
  \caption{(a) Model training with prosody learning mechanism, where a prosody learner 
  is designed to model the prosody information from mel-spectrum of speech in training dataset. 
  (b) Prosody predictor, where a pre-training language model is introduced to 
  improve the prosody prediction results. }
  \label{TTS-architecture}
\end{figure*}

\subsection{Local Attention Block} 

\subsubsection{Self-Attention of Transformer} 

Recent years, the self-attention mechanism is proposed to 
capture the context-sensitive features in Transformer 
\cite{AttentionAllYouNeed}, which is widely used in 
various nature language process task 
\cite{Bert,Xlnet,BioBERT,recent-trends-nlp}. 
Due to the similarity of machine translation and speech synthesis, 
the structure of Transformer is also applied in TTS systems, 
such as Transformer TTS \cite{TransformerTTS}, FastSpeech 
\cite{FastSpeech} and AlignTTS \cite{AlignTTS}. 
Although the self-attention structure has been proven to 
achieve satisfactory performance in speech synthesis, 
it is not perfect for TTS system. In detail, 
the self-attention is designed to model the dependence of 
any two elements in the input sequence, which does not 
make good use of the law of the influence of the input 
sequence position distance relationship on pronunciation. 
On the other hand, the positional encodings are used to 
inject the positional information of input sequence, 
but it also limits the maximum length of the input sequence.

\subsubsection{Local Attention}

In our works, we design a novel self-attention network, 
named as local attention, which is more suitable for 
TTS systems. Let $ \boldsymbol{h} = \{ h_1, h_2, ..., h_n \} $ 
denote the input sequence, where each element in sequence is vector. 
The attention function is described as mapping a query and 
a set of key-value pairs to an output. Following the Transformer, 
the query sequence, the key sequence and the value sequence 
are calculated by
\begin{align} 
  \boldsymbol{q}, \boldsymbol{k}, \boldsymbol{v} &= 
          \boldsymbol{h} \mathbf{W}^q, 
          \boldsymbol{h} \mathbf{W}^k, 
          \boldsymbol{h} \mathbf{W}^v 
\end{align}

Different from Transformer using the dot products of $\boldsymbol{q}$ 
and $\boldsymbol{k}$ to compute the attention weights, 
the local attention introduces a relative position matrix 
to model the positional relationship between elements in 
$\boldsymbol{q}$ and $\boldsymbol{k}$. 
The attention weights before softmax function can be calculated by
\begin{align} 
  A_{i,j} = \left\{ \begin{matrix}
    \left( q_i \right )^{\top} \mathbf{W}_{(i-j)}^{\text{loc}} k_j & | i-j | \le T \\
    - inf &  |i-j| > T
  \end{matrix} \right.
\end{align}
where $\mathbf{W}^{\text{loc}} = \{ \mathbf{W}_{-T}^{\text{loc}}, ..., 
\mathbf{W}_{0}^{\text{loc}}, ..., \mathbf{W}_{T}^{\text{loc}} \}$ 
denotes the relative position matrices, and $T$ denotes the maximum 
relative position distance considered by the attention calculation. 
And the local attention can be written as
\begin{align} 
  \text{Local-Attention} (\boldsymbol{h}) &= \text{Masked-Softmax}( \frac{ \mathbf{A} }{ \sqrt{d_k} } ) \boldsymbol{v}
\end{align}
where $ \mathbf{A} = \{ A_{i,j} \}_{n \times n} $, and $ \sqrt{d_k} $ is the scaling factor.

\subsubsection{Network Structure} 

Similar to Transformer, the multi-head mechanism is also 
adopted in the local attention. In addition, the local 
attention block is designed to facilitate multi-layer 
stacking, which is composed of a multi-head local attention 
and a 2-layer 1D convolution network. The residual 
connection \cite{resnet}, layer normalization \cite{LayerNorm} 
and dropout \cite{Dropout} are also used, same as Transformer. 
Note that, in the local attention, the relative positional 
information can be modeled by the relative position matrices 
$\mathbf{W}^{\text{loc}}$, thus the positional encodings 
can be removed. In our opinion, compared with the positional 
encodings, the relative position matrices make the network 
more capable of modeling the location information.

\subsection{Prosody Modeling} 

\subsubsection{Prosody Learning} 

The prosody of speech affects the listening sensation 
of the entire speech, especially for long text speech. 
To obtain synthetic speech that approximates a real person, 
prosody information must be modeled when synthesizing speech 
of long text. Meanwhile, using different 
prosodies to speak, the same text will obtain different speech, thus it is 
necessary for advanced TTS systems to consider the effect of 
prosody on pronunciation. 

In our works, we propose the prosody learning mechanism to 
model the prosody of speech in training. As shown in Figure \ref{figure-prosody-modeling}, 
a prosody learner is designed to calculate the prosody embedding 
according to the mel-spectrum, which is added into phoneme 
embedding to affect the training of TTS systems. In detail, 
we first calculate the alignment between the phoneme sequence 
and the mel-spectrum sequence using the alignment loss 
introduced in AlignTTS \cite{AlignTTS}. The mel-spectrum 
sequence is passed through multi-layer 1D convolutional network, 
where residual connection and relu activation function are used, 
and then it is pooled according to the alignment between phoneme 
and mel-spectrum to output a prosody representation sequence. 
The prosody representation sequence is low-dimensional 
(for example, 3), of which length is equal to the length of 
the phoneme sequence. In addition, a prosody mapping network, 
just a linear layer, is designed to generate prosody embedding 
from the prosody representation sequence. The prosody embedding 
is added into the phoneme embedding in the mel-spectrum predictor 
and the duration predictor.

\subsubsection{Prosody Prediction}

In our daily speech, the prosody of each phoneme is often more 
corresponding to the meaning of the word. Therefore, in order to 
predict more satisfactory prosody, we introduce the pre-trained 
language model to design a prosody predictor. 
As shown in Figure \ref{figure-prosody-predictor}, the prosody predictor is composed of 
the phoneme embedding layer, the pre-trained language model, 
and multiple local attention blocks and linear layers. 
The word embeddings from the pre-trained language model is 
first upsampled to align with the phoneme sequence. 
The phoneme embedding is passed through 3-layer 1D convolution 
networks, and the is added to the upsampled word embedding.
Multiple stacked local attention blocks are used to model 
the dependencies between sequence elements. And the last 
linear layer outputs the distribution of the prosody representation. 
Considering the variability of prosody, the mix density network \cite{MDNs} 
is used to model the distribution of the prosody representation. 
In detail, the last linear layer outputs the mean, variance and 
weight of multiple mixed Gaussian distributions for each element 
of the prosody representation sequence.

In order to train the prosody predictor, we need first finish 
the training of the prosody learner, and use it to generate 
the prosody representation sequence of speech in training dataset. 
The generated prosody representation sequence is used 
as the target to guide the training of the prosody predictor. 

\subsection{Text-to-Speech System }

\subsubsection{Architecture} 

Using the local attention and the prosody modeling, we design 
a feed-forward TTS system based on AlignTTS. 
As shown in Figure \ref{TTS-architecture}, 
the proposed system consists of a mel-spectrum predictor for 
transforming the phoneme sequence into the mel-spectrum, 
a duration predictor for predicting the duration of each phoneme, 
a prosody learner for modeling the prosody of speech, 
and a prosody predictor for predicting the prosody in inference. 
The mel-spectrum predictor and the duration predictor are 
modification version of AlignTTS, where the FFT block of 
AlignTTS is replaced by the local attention block and 
the positional encoding is removed. The length regulator 
is the same as that in AlignTTS, which requires correct 
alignment between phonemes and mel-spectrum in training and 
the phoneme durations predicted by the duration predictor in inference.

\subsubsection{Training}
In training, we first model and extract the alignment between 
phonemes and mel-spectrum according to \cite{AlignTTS}. 
The mel-spectrum predictor, the duration predictor and 
the prosody learner are trained together, where the extracted 
alignment is used to guide the average pooling in prosody learner 
and converted into the duration sequence as the target of the 
duration predictor. After finishing their training, the prosody 
representation sequence of each speech in training dataset is 
calculated by the trained prosody learner, which is used as 
the target to guide the training of the prosody predictor.

\subsubsection{Inference} 

In inference, according to the text and its phonemes, the prosody 
predictor first predicts the distributions of the prosody representation. 
The prosody representation sequence sampled from the predicted 
distributions is passed through the prosody mapping network to 
generate the prosody embedding that is added into the phoneme 
embedding to predict the duration sequence in the duration predictor. 
Based on the prosody embedding and the duration sequence, 
the mel-spectrum predictor generates the mel-spectrum. 
Finally, A neural vocoder, such as WaveGlow \cite{WaveGlow}, 
is used to generate the waveform according to the predicted mel-spectrum.

\section{Experiments} 

\subsection{Dataset} 

We test the performance of our model in synthesizing English and Mandarin. 
The LJSpeech dataset \cite{ljspeech17} is chosen as the English speech 
dataset, which consists of 13100 short audio clips of a single speaker 
and corresponding transcriptions. The Mandarin speech dataset uses
the Chinese Standard Mandarin Speech Copus from 
Databaker \cite{Biaobei}, 
which is composed of 10000 speech of a 
female and transcriptions. These datasets are randomly divided into 
2 sets: $ 98\% $ of samples for training and the rest for testing. 
The sample rate of audio is set to 22050 Hz. The mel-spectrums are 
computed through a short-time Fourier transform with Hann windowing, 
where 1024 for FFT size, 1024 for window size and 256 for hop size. 
The STFT magnitude is transformed to the mel scale using 80 channel 
mel filter bank spanning 60 Hz to 7.6kHz.

For English synthesis, the characters of text are directly used as 
the input of model, and for Mandarin synthesis, the pinyin 
sequence of text is used as input. In addition, 
in order to verify the prosody modeling capabilities, 
all marks of rhythm boundaries are replaced by spaces in our experiences. 

\subsection{Model Configuration}

In our experiments, the number of the local attention block is set 
to 6 in the alignment learner. The mel-spectrum predictor contains 
6 local attention blocks on both the phoneme side and the mel-spectrum 
side, and the duration predictor includes 3 local attention blocks. 
The channel of these network is all set to 768. In all local attention 
blocks, the number of attention head is set to 2, the kernel size of 
1D convolution is set to 3, and the maximum relative position distance 
of the relative position matrices is set to 10. The number of the 1D 
convolution layers is 4 in the prosody learner, and the number of the 
local attention blocks is also 4 in the prosody predictor. In addition, 
AlBert \cite{Albert} is chosen as the pre-trained language model, and 
we directly use the official open-source trained model in our experiments. 

Our TTS system is trained on 2 NVIDIA V100 GPUs with batch size of 16 
samples on each GPU. The Adam optimizer \cite{Adam} with $ \beta_1 = 0.9 $, 
$ \beta_2 = 0.98 $, $ \varepsilon = 10^{-9} $ is used in our case. 
In addition, we adopt the same learning rate schedule described in 
\cite{AttentionAllYouNeed} in whole training.


\subsection{Results}
In order to evaluate the performance of our model, we compare 
the quality of speech synthesized by Tacotron 2 \cite{Tacotron2}, 
FastSpeech \cite{FastSpeech}, AlignTTS \cite{AlignTTS} and our model. 
In details, the text transcription in test datasets is used as 
the input of these model to predict the mel-spectrum, and then 
the WaveGlow vocoder is used to generate the waveform from the 
predicted mel-spectrum. The audios synthesized by different models 
are rated together with the ground truth audio (GT) by 50 testers. 
And then the mean opinion score (MOS) is calculated as the evaluation 
indicator.

The evaluation results are shown in Table \ref{Table-TTS}, where the synthetic 
speech of English and Mandarin are rated separately. It can be 
seen that our model gets the same performance as AlignTTS, and 
outperforms Tacotron 2 by 0.09 in synthesizing English. 
In Mandarin experiments, our model obtains significantly improved 
performance than AlignTTS. According to our analysis, the synthetic 
speech from our model has more satisfactory prosody, which makes 
the overall listening experiences of speech greatly improved, 
especially for Mandarin.


\begin{table}[t]
  \begin{center}
  \caption{The comparison of MOS among different TTS systems.}
  \begin{tabular}{p{2cm}p{2.5cm}p{2cm}<{\centering}}
  \toprule
  \textbf{Language}&\textbf{Method}&\textbf{MOS} \\
  \midrule
  \multirow{5}*{English} 
        &  Ground Truth  &  $ 4.56 \pm 0.05 $  \\
        &  Tacotron 2    &  $ 3.96 \pm 0.13 $   \\
        &  FastSpeech    &  $ 3.88 \pm 0.11 $  \\
        &  AlignTTS      &  $ 4.05 \pm 0.12 $  \\
        &  \textbf{Ours} &  $ \bm{ 4.05 \pm 0.06 } $  \\
  \midrule
  \multirow{5}*{Mandarin} 
        &  Ground Truth  &  $ 4.68 \pm 0.05$  \\
        &  Tacotron 2    &  $ 3.94 \pm 0.10 $   \\
        &  FastSpeech    &  $ 3.81 \pm 0.08 $  \\
        &  AlignTTS      &  $ 3.95 \pm 0.11 $  \\
        &  \textbf{Ours} &  $ \bm{4.03 \pm 0.09} $  \\
  \bottomrule
  \end{tabular}
  \label{Table-TTS}
  \end{center}
\end{table}

\subsection{Ablations}


\subsubsection{Pre-trained Language Model}
In our model, the pre-trained language model is introduced to 
predict the prosody information. In order to verify the effectiveness, 
we conduct a comparative experiment on whether the prosody predictor 
needs information from the pre-trained language model. A comparative 
prosody predictor without the pre-trained language model is trained 
and predict the prosody presentation for synthesizing speech. These 
speech predicted by different prosody predictor are rated together, 
and the results are shown in Table \ref{Table-pre-trained}. We can see that the introduction 
of the pre-trained language model can significantly improve the quality 
of synthetic speech, especially in prosody. Meanwhile, 
even without the pre-trained language model in prosody predictor, 
our model can still obtain better performance than AlignTTS, which 
verifies that the separation of prosody information is also beneficial 
to the mel-spectrum modeling.

\begin{table}[t]
  \begin{center}
  \caption{The effect of the pre-trained language model on Mandarin synthesis.}
  \begin{tabular}{p{5cm}p{2cm}<{\centering}}
  \toprule
  \textbf{Method} & \textbf{MOS} \\
  \midrule
  AlignTTS                               &  $ 3.95 \pm 0.11 $  \\
  Without Pre-trained Language Model     &  $ 3.97 \pm 0.11 $  \\
  With Pre-trained Language Model        &  $ 4.03 \pm 0.09 $  \\
  \bottomrule
  \end{tabular}
  \label{Table-pre-trained}
  \end{center}
\end{table}

\begin{table}[t]
  \begin{center}
  \caption{The effect of the dimension of the prosody representation on Mandarin synthesis.}
  \begin{tabular}{p{3cm}p{2cm}<{\centering}}
  \toprule
  \textbf{Dimension} & \textbf{MOS} \\
  \midrule
  1              &  $ 3.98 \pm 0.12 $  \\
  3              &  $ 4.03 \pm 0.09 $  \\
  6              &  $ 4.02 \pm 0.05 $  \\
  10             &  $ 3.99 \pm 0.13 $  \\
  \bottomrule
  \end{tabular}
  \label{Table-dimension}
  \end{center}
\end{table}

\subsubsection{Dimension of Prosody Representation}
In order to model the prosody information, we introduce 
the prosody representation that is learned from speech in 
training process and is predicted by the prosody predictor in 
inference process. The prosody representation is a compressed 
representation sequence of speech, which can be used to 
reconstruct speech when providing the phoneme sequence. 
Therefore, the dimension of the prosody representation is 
a significant hyperparameter, since too high dimension will 
introduce too much information from the speech, which affects 
the modeling of the pronunciation of each phoneme in the 
mel-spectrum predictor. We experimentally analyze the influence 
of this dimension on the performance of TTS system. 
The synthetic speech from different TTS systems with 
different dimensions of the prosody representation are 
rated by our tester together. The results are shown in Table \ref{Table-dimension}. 
It can be seen that appropriate dimension can make the model get 
the best performance, and too high dimension reduces model performance.


\section{Conclusion}

Based on the local attention, a feed-forward text-to-speech 
system without the limitation of input text length is designed. 
Meanwhile, the prosody learning mechanism is proposed to model 
the prosody of speech, where the prosody information is learned 
from speech by a prosody learner in training process. 
In order to predict more satisfactory prosody in inference, 
the pre-trained language model is used to introduce the semantic 
feature. Experiments on English synthesis and Mandarin synthesis 
show that a significant improvement in the prosody of speech has 
been obtained in our proposed TTS systems.

\section{Acknowledgements}

This paper is supported by National Key Research and Development Program of China under grant No. 2018YFB1003500, No. 2018YFB0204400 and No. 2017YFB1401202. 
Corresponding author is Jianzong Wang from Ping An Technology (Shenzhen) Co., Ltd.

\bibliographystyle{IEEEtran}

\bibliography{references}


\end{document}